# Towards a Novel Privacy-Preserving Distributed Multiparty Data Outsourcing Scheme for Cloud Computing with Quantum Key Distribution

D. Dhinakaran*[1], D. Selvaraj[2], N. Dharini [3], S. Edwin Raja[4], C. Sakthi Lakshmi Priya[5]

**Abstract:** The intersection of cloud computing, blockchain technology, and the impending era of quantum computing presents a critical juncture for data security. This research addresses the escalating vulnerabilities by proposing a comprehensive framework that integrates Quantum Key Distribution (QKD), CRYSTALS-Kyber, and Zero-Knowledge Proofs (ZKPs) for securing data in cloud-based blockchain systems. The primary objective is to fortify data against quantum threats through the implementation of QKD, a quantum-safe cryptographic protocol. We leverage the lattice-based cryptographic mechanism, CRYSTALS-Kyber, known for its resilience against quantum attacks. Additionally, ZKPs are introduced to enhance data privacy and verification processes within the cloud and blockchain environment. A significant focus of this research is the performance evaluation of the proposed framework. Rigorous analyses encompass encryption and decryption processes, quantum key generation rates, and overall system efficiency. Practical implications are scrutinized, considering factors such as file size, response time, and computational overhead. The evaluation sheds light on the framework's viability in real-world cloud environments, emphasizing its efficiency in mitigating quantum threats. The findings contribute a robust quantum-safe and ZKP-integrated security framework tailored for cloud-based blockchain storage. By addressing critical gaps in theoretical advancements, this research offers practical insights for organizations seeking to secure their data against quantum threats. The framework's efficiency and scalability underscore its practical feasibility, serving as a guide for implementing enhanced data security in the evolving landscape of quantum computing and blockchain integration within cloud environments.

## 1. Introduction

The introduction of cloud computing has completely changed how businesses handle, store, and use their data in the current era of exponential data expansion and digital transformation [1]. The scalability, cost-efficiency, and flexibility of cloud services have made them indispensable for businesses across various sectors. However, as the volume of sensitive and confidential data migrated to the cloud continues to surge, data security as well as privacy have become paramount concerns [2-4]. Protecting sensitive information from malicious actors and ensuring the confidentiality of data in the cloud are formidable challenges. Traditionally, cryptographic techniques such as public, symmetric-key cryptography, and hash functions have been employed to secure data during transit and at rest [5]. These methods, while effective in classical computing environments, face an imminent threat with the rise of quantum computing. The computational power of quantum computers poses a significant risk to the security of existing cryptographic protocols, making it crucial for researchers and practitioners to explore innovative solutions that can withstand quantum attacks.

QKD has evolved as a ground-breaking method to strengthen data security in the context of cloud computing (CC) in answer to this urgent requirement. By transferring quantum keys, QKD uses the ideas of quantum physics to create reliable communication channels [6-9]. Unlike classical cryptographic methods, QKD is immune to quantum attacks, providing an unprecedented level of security. This research article explores the integration of quantum key distribution into cloud computing environments as a quantum-safe solution to enhance data security [10]. We delve into the theoretical foundations of QKD and its applicability in cloud security, addressing the unique challenges and opportunities it presents [11-13]. We also provide a comprehensive analysis of the performance and practical aspects of implementing QKD in the cloud, examining key metrics such as encryption and decryption

[1]*Department of Computer Science and Engineering,*
*Vel Tech Rangarajan Dr. Sagunthala R&D Institute of Science and Technology, Chennai, India*
*ORCID ID :  0000-0002-3183-576X*

[2] *Department of Electronics and Communication Engineering, Panimalar Engineering College, Chennai, India.*
*ORCID ID :  0000-0001-5486-5390*

[3]*Department of Computer Science and Engineering (Cyber Security),*
*R.M.K College of Engineering and Technology, Chennai, India*
*ORCID ID :  0000-0002-9459-4019*

[4] *Department of Computer Science and Engineering,*
*Vel Tech Rangarajan Dr. Sagunthala R&D Institute of Science and Technology, Chennai, India*
*ORCID ID :  0000-0002-2948-9669*

[5] *Department of Computer Science and Engineering,*
*P.S.R Engineering College, Sivakasi, India*
*ORCID ID :  0000-0003-4347-3732*
*\* Corresponding Author Email: dhinaads@gmail.com*

times, quantum key generation rates, and scalability.

## 1.1. Objectives

*Explore Quantum-Safe Blockchain-Cloud Security:* The primary objective of this research is to thoroughly explore the principles of Quantum Key Distribution (QKD) and its integration with blockchain storage processes in cloud computing. We aim to understand the core concepts and mechanisms of QKD and how it can be harnessed to secure data stored on blockchain platforms in the cloud.

*Mitigate Quantum Threats to Blockchain Data:* The research seeks to address the quantum threats posed to data stored on blockchain platforms. By proposing and implementing a quantum-safe framework that combines QKD and CRYSTALS-Kyber with blockchain storage, we provide a robust solution that ensures the integrity and confidentiality of data while utilizing blockchain's inherent features.

*Performance Analysis and Scalability for Blockchain-Cloud Integration:* We conduct a comprehensive performance analysis to evaluate the proposed quantum-safe framework's impact on encryption and decryption times, quantum key generation rates, and scalability. This analysis is particularly relevant in the context of blockchain storage processes and cloud operations, providing insights into the practicality and efficiency of the integration.

*Innovative Application of Zero-Knowledge Proofs (ZKPs):* The research extends its focus to the integration of Zero-Knowledge Proofs (ZKPs) within cloud security. Our objective is to explore the innovative application of ZKPs in combination with quantum-safe measures, ensuring data privacy and confidentiality while leveraging the principles of quantum physics.

*Practical Implementation Guidance:* We provide practical guidance and implementation strategies for organizations aiming to adopt quantum-safe security and ZKP-based verification in their blockchain storage and cloud computing operations. This objective facilitates the real-world adoption of advanced security measures while maintaining the scalability and efficiency required for blockchain-based cloud applications.

*Strategic Adaptation to Emerging Technologies:* The research outlines strategic directions for organizations to adapt to evolving quantum technologies, blockchain storage processes, and the integration of ZKPs. By staying ahead of emerging technological trends, organizations can remain resilient to quantum threats and leverage innovative solutions for secure cloud-based blockchain operations.

The objectives underscore the importance of addressing the quantum threat to data security in cloud computing (CC)also seeks to provide practical solutions that safeguard data confidentiality, integrity, and availability in an ever-evolving cloud landscape.

## 1.2. Contributions

*Quantum-Safe Cloud Security Framework:* This research introduces a groundbreaking quantum-safe framework that leverages Quantum Key Distribution (QKD) in the context of blockchain storage processes within cloud computing. The framework offers a quantum-resistant solution for securing data stored on blockchain platforms in cloud environments, ensuring the integrity and confidentiality of sensitive information.

*Protection Against Quantum Attacks and Blockchain Threats:* In addition to addressing quantum attacks, the research explores the intersection of quantum threats and blockchain storage. By proposing the integration of QKD with blockchain processes in the cloud, it provides a comprehensive approach to safeguarding blockchain data from quantum adversaries. This contribution bridges the gap between quantum-safe security and the blockchain ecosystem.

*Performance Analysis and Scalability in Blockchain-Cloud Integration:* A rigorous performance analysis is conducted to evaluate the quantum-safe framework's impact on encryption and decryption times, quantum key generation rates, and scalability, particularly in the context of CRYSTALS-Kyber and blockchain storage. This analysis sheds light on the efficiency and practicality of integrating QKD with blockchain processes in cloud environments.

*Innovative Approach to Zero-Knowledge Proofs (ZKPs) in Cloud Security:* The research extends its innovative approach to the realm of Zero-Knowledge Proofs (ZKPs) within cloud security. By integrating ZKPs with quantum-safe cloud security, the research introduces a pioneering concept that enhances data privacy and confidentiality while leveraging the principles of quantum physics.

*Practical Implementation Guidance for Blockchain and Cloud:* Practical guidance and implementation strategies are provided for organizations seeking to implement quantum-safe and ZKP-based security in blockchain storage and cloud computing. This contribution empowers organizations to protect and verify the integrity of their data while maintaining the scalability and efficiency required for cloud-based blockchain applications.

*Strategic Adaptation to Evolving Technologies:* The research not only addresses current security challenges but also outlines strategies for organizations to adapt to evolving quantum technologies and blockchain storage processes. By staying ahead of the curve, organizations can remain resilient to quantum threats and leverage emerging technologies for secure cloud-based blockchain operations.

With the ever-growing need for robust data security and the looming threat of quantum attacks, this research seeks to pave the way for a new era of quantum-safe cloud computing, where data confidentiality and integrity remain inviolable. The rest of this article is organized as follows. Section 2 offers a literature survey by addressing various researchers who provided the solution to the problem statement. Section 2 also discusses the current state of data security in CC and the vulnerabilities that quantum computing introduces. In Section 3, we present our proposed approach for implementing QKD in cloud environments, offering a novel solution for quantum-safe cloud security. Section 4 delves into our experimental methodology and the outcomes of our performance analysis, demonstrating the viability of our approach. In Section 5, we conclude by summarizing our results and suggesting possible avenues for further investigation.

## 2. Literature Survey

The ever-evolving landscape of cloud computing and blockchain technology has brought forth new opportunities and challenges in data security and privacy. As organizations increasingly rely on cloud environments for data storage and processing, the need to protect sensitive information has become paramount. The integration of blockchain storage processes into cloud computing environments presents new opportunities for data management. However, it also introduces security challenges, particularly in the context of the looming threat of quantum computing. Current cryptographic methods face vulnerabilities to quantum attacks, jeopardizing the confidentiality as well as integrity of data stored on blockchain platforms in the cloud. Moreover, ensuring data privacy as well as compliance with evolving regulations remains a complex task. This literature survey explores existing research and developments in the domains of quantum-safe security, blockchain storage, CRYSTALS-Kyber, and Zero-Knowledge Proofs (ZKPs) within the framework of cloud computing. It seeks to provide a comprehensive overview of the state of the art, highlight recent approaches, and identify areas where current solutions fall short of achieving the overarching goal of robust data security in cloud-based blockchain storage.

The foundational work of Gisin et al. [14] delves into the principles of quantum cryptography and the potential applications of Quantum Key Distribution (QKD) to secure data communication. QKD, grounded in the laws of quantum mechanics, offers a quantum-safe approach to data security. Wanget al. [15] innovatively combines homomorphic authenticator through random masking to establish a public cloud data auditing system. This approach satisfies various requirements. Additionally, the incorporation of bilinear aggregate signatures enables efficient management of multiple auditing tasks, extending the system's applicability to a multi-user environment. This advancement allows a Third-Party Auditor (TPA) to conduct simultaneous auditing tasks with enhanced efficiency and privacy preservation.

Ogielaet al. [16] introduce a novel approach to data security by combining cryptographic threshold techniques with linguistic methods for describing shared secrets, resulting in intelligent linguistic threshold schemes. This innovative protocol class is designed for securing data across different levels of management within entities, considering variations in both structure and environment. The paper particularly focuses on applying these solutions to service management and security at various data management levels, with a special emphasis on assessing the feasibility of implementing these protocols in cloud management processes.

Safaret al. [17] aims to pinpoint and analyze challenges related to cloud computing, with a specific emphasis on data security. The objective includes conducting a comprehensive scientific review and comparative analysis of recent research studies. Kumar [18] integrates DNA-based algorithms with the AES algorithm to establish data security platform. This innovative approach employs DNA cryptography technology as well as the AES method. The combined use of approach not only enhances data security but also presents a technologically robust solution for improving cloud security. Xiaoyu et al. [19] introduces a novel dynamic hash authentication scheme utilizing Merkle trees. The proposed trust empowerment management system is mapped onto a user dimension (letters) and a management technology side. Emphasis is placed on data encryption technology, offering comprehensive cloud storage security solutions covering data design and demotion. Furthermore, the paper addresses issues in information security standards by suggesting a fusion of cloud security standards with an evaluation system.

Namasudra et al. [20] proposes an access control model for efficient and secure cloud computing. The model employs ABE, DHT network, and IDTRE. Initially, data are encrypted based on user attributes, leading to encapsulated and extracted ciphertexts. The IDTRE algorithm encrypts the decryption key, combining it with the ciphertext to form ciphertext shares. The ciphertext shares are then distributed across the hash table network, while encapsulated ciphertexts are stored on cloud servers, ensuring a robust framework for secure access control in the CC environment.

The problem of enabling a third-party auditor to confirm the accuracy of rapidly changing information stored in the public cloud on behalf of the customer is addressed by Q. Wang et al. [21]. This approach eliminates client involvement, enhancing efficiency in cloud computing. The focus is on supporting data dynamics, insertion, as well as deletion, making the system more practical for a broader range of cloud services. The paper distinguishes itself by achieving both public auditability and support for dynamic data operations, overcoming limitations found in prior works. The proposed verification scheme seamlessly integrates these features, with improvements to existing proof of storage models.

Kaur et al. [22] focuses on addressing challenges in cloud computing, particularly emphasizing data privacy concerns. Cloud computing, as a virtual pool of resources delivered via the internet, aims to resolve daily computing problems. The primary issues under consideration include data privacy, security, anonymity, and reliability. The research proposes a work plan centered on enhancing security in the cloud, with a primary emphasis on data privacy. This is to be achieved through the implementation of encryption algorithms, addressing the pivotal concern of security assurance by cloud providers from various perspectives of cloud customers. The popularity of cloud computing for cost-effective storage and computing services in organizations comes with challenges related to data confidentiality, integrity, and access control. While existing approaches address these concerns, some limitations persist, including risks of collusion attacks and computational inefficiency due to a high number of keys. To overcome these issues, Saroj et al. [23] propose a scheme employing threshold cryptography. With this method, the data owner groups users and gives a single key for decryption to each group, with each group member possessing a portion of the key. Access control is managed using a capability list. This scheme enhances data confidentiality robustly and concurrently reduces the number of keys required.

Ahmad Khan et al. [24] employs Elliptic Curve Cryptography (ECC) for data encryption in the cloud environment, leveraging its small key size. ECC's compact key size not only enhances computational efficiency but also minimizes energy consumption. The research demonstrates that ECC is a fast and efficient choice for data protection in cloud computing, highlighting its ability to reduce computational power and increase overall efficiency in securing data within the cloud environment. Zhang et al. [25] introduces a match-then-decrypt technique. In order to determine whether the characteristic of the private key meets the concealed access policy without requiring full decryption, this method computes particular ciphertext components. Special attribute secret key elements are developed to allow fast decryption by enabling pairing aggregation through the stage of decryption. The study first offers an anonymous Attribute-Based Encryption (ABE) design that forms the foundation. Next, a security-enhanced version is built by adding one-time signatures that are strongly existentially unforgeable.

Ranjan et al. [26] ensures secure communication and information concealment from unauthorized users. The model combines RSA encryption and digital signature techniques, making it compatible with various cloud computing features. To be more precise, the MD5 algorithm is used for authentication, and the RSA encryption algorithm is used for data secrecy. The integration of these cryptographic techniques enhances overall security in the cloud computing environment. Blockchain technology, best known for its role in cryptocurrencies, holds the promise of providing secure and immutable data storage. Zheng et al. [27] discuss the architecture and consensus mechanisms of blockchain technology. These fundamentals are vital for comprehending blockchain storage in cloud applications, where data immutability is crucial. Shi et al. [28] discuss techniques for enhancing storage security in cloud computing, they address the privacy-preserving challenge of determining the maximum value across several secrets, the author adopted the quantum protocol.

The survey reveals several common issues encountered by existing approaches. Many studies identify concerns related to data privacy, authentication, and data integrity in cloud environments. Some challenges include potential vulnerabilities in traditional cryptographic methods and the lack of robust solutions for emerging technologies like quantum computing. Our proposed framework addressing the intersection of cloud computing, blockchain, and quantum computing provides a comprehensive response to these challenges. Unlike certain existing approaches that may focus on individual aspects, our framework integrates Quantum Key Distribution (QKD), CRYSTALS-Kyber, and Zero-Knowledge Proofs (ZKPs). This approach addresses the limitations of conventional cryptographic methods and enhances the security of cloud-based blockchain systems. One significant contribution of our work is the emphasis on quantum-safe protocols. While some existing approaches may not adequately prepare for the imminent era of quantum computing, our framework leverages QKD to fortify data against potential quantum threats. The use of CRYSTALS-Kyber, a resilient lattice-based cryptographic mechanism, further strengthens security measures, providing a more robust solution compared to traditional approaches. Additionally, our work stands out by conducting a rigorous performance evaluation, considering factors such as encryption and decryption processes, quantum key generation rates, and overall system efficiency. This detailed analysis, including practical implications like file size, response time, and computational overhead, positions our framework as a viable and efficient solution for addressing the identified challenges in secure cloud computing.

## 3. Proposed System Model

Our proposed work presents a holistic approach to privacy-preserving distributed multiparty data outsourcing in cloud computing. It utilizes a blockchain-based trusted authority system, post-quantum cryptography, and Zero-Knowledge Proofs to enhance security, privacy, and efficiency in key management and data processing. This ground-breaking architecture has the potential to completely change how key pairs are created and disseminated while providing a revolutionary response to the problems of data security as well as privacy in the age of CC.

**3.1. Motivations**

*Quantum Computing Threat to Blockchain Data:* The growing threat posed by quantum computing to data security in blockchain storage is a primary motivation for this research. Quantum computers have the potential to compromise classical cryptographic methods used in blockchain platforms. This research is driven by the urgency to address this quantum threat and protect the integrity as well as confidentiality of blockchain-stored data.

*Security and Privacy in Blockchain-Cloud Integration:* As blockchain technology becomes increasingly integrated into CC, ensuring the security as well as privacy of data becomes paramount. Motivated by the need to safeguard sensitive data on blockchain platforms within cloud environments, this research explores innovative quantum-safe and ZKP-based solutions to enhance security.

*Blockchain Compliance and Data Protection:* Data protection regulations and compliance requirements continue to evolve, demanding the implementation of robust security measures. This research is motivated by the need to assist organizations in achieving compliance while ensuring data security in blockchain storage processes within the cloud.

*Cybersecurity Threats:* The rise in cyberattacks and security incidents necessitates proactive measures. The research is motivated by the urgency to defend against evolving threats and enhance the resilience of blockchain storage and cloud systems against malicious actors.

*Cutting-Edge Cloud and Blockchain Technologies:* The dynamic evolution of cloud and blockchain technologies presents both challenges and opportunities. Motivated by the desire to stay at the forefront of security in these domains, this research explores innovative approaches to quantum-safe and ZKP-based data security.

*Innovation in Quantum-Safe Blockchain Security:* The research's motivation extends to pioneering new horizons in blockchain security. By integrating Quantum Key Distribution, CRYSTALS-Kyber, and Zero-Knowledge Proofs, this research introduces groundbreaking methods that harness quantum principles and blockchain technology to ensure data confidentiality and integrity.

*Practical Guidance for Implementation:* This research is motivated by the practical need to provide clear guidance and strategies for organizations looking to implement quantum-safe and ZKP-based security in their blockchain-cloud ecosystems. The objective is to make advanced security practices accessible and actionable for a wide range of organizations.

The six phases of the suggested framework are as follows:

*1. Key Generation:* The trusted authority generates unique key pairs for system users. Private keys are securely stored, and public keys are distributed.

*2. Data Encryption:* System users encrypt their data using the public key of the trusted authority.

*3. Blockchain Storage:* The encrypted data is stored on the blockchain, which ensures transparency and security.

*4. Decryption Process:* When a system user needs to decrypt the data, they request the private key from the trusted authority.

*5. Private Key Verification:* The trusted authority verifies the user's identity before sending the private key.

*6. Data Decryption:* The system user can then decrypt the data using the provided private key.

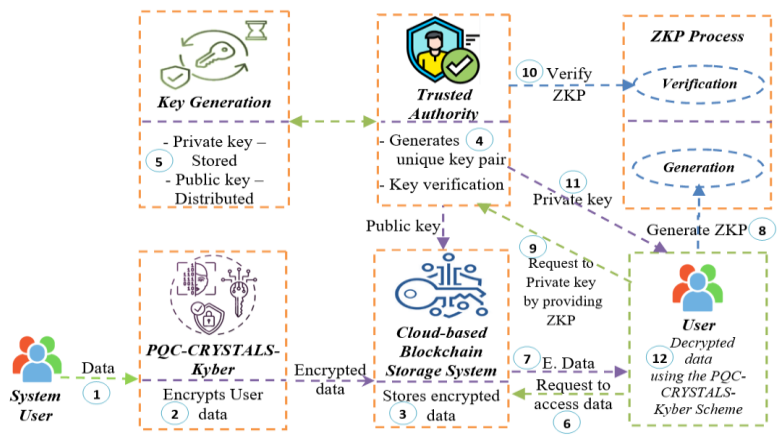

**Fig. 1.** Proposed privacy-preserving Framework

In our proposed work, we introduce a novel and innovative framework for privacy-preserving distributed multiparty data outsourcing in the framework of cloud computing. The central pillar of our approach is a blockchain-based trusted authority system that reimagines the conventional methods of key generation and distribution [29-31]. This revolutionary system addresses the evolving challenges of data security and privacy in an era where cloud-based data processing has become the norm. The Trusted Authority (TA) serves as the linchpin of our framework, playing a pivotal role in the secure and efficient distribution of cryptographic keys to System Users (SUs) as shown in Fig. 1. A distinct key pair, made up of a private key as well as a public key, is given to every SU. The private key is carefully guarded by the TA, ensuring the utmost confidentiality, while the public key is made available on the blockchain, a distributed and immutable ledger.

This blockchain-based approach has a transformative impact on key management [32]. By utilizing the blockchain, we achieve unparalleled levels of transparency and security in the distribution process. The SUs can effortlessly verify the authenticity of the TA's public key, enhancing trust and accountability. Furthermore, all key-related transactions are recorded on the blockchain, creating an unforgeable record of events that is resistant to tampering [33]. To safeguard the privacy and security of the data, SUs employ post-quantum cryptography (PQC) - CRYSTALS-Kyber for data encryption. This advanced cryptographic technique ensures that sensitive information remains confidential during its transmission and storage in the cloud server, which is semi-trusted. A key innovation in our approach is the utilization of Zero-Knowledge Proofs (ZKPs) for private key verification. ZKPs are cryptographic protocols that enable an SU to convince the TA that they possess knowledge of their private key without revealing the key itself. This process is integral to our scheme, offering an extra layer of security and privacy. When an SU needs to decrypt their data, they request their private key from the TA. Prior to releasing the private key, the TA requests a ZKP from the SU to confirm their knowledge of the key. This allows the TA to verify the SU's identity without ever accessing the private key.

ZKPs provide several advantages over traditional identity verification methods. They enhance security by ensuring that the private key is never revealed to the TA, preventing any unauthorized access to sensitive information. Moreover, ZKPs preserve user privacy, as the TA only learns that the SU knows their private key, not the key itself. This approach is highly convenient for users, as they are not required to provide additional personal information to the TA, simplifying the identity verification process. As part of our work, we have devised a comprehensive example of how this ZKP-based identity verification process works. An SU first generates a ZKP that attests to their knowledge of the private key. They then send this ZKP to the TA, initiating the identity verification process. The TA meticulously verifies the ZKP, ensuring its validity. If the ZKP is confirmed as valid, the TA proceeds to release the private key to the SU, allowing them to decrypt their data. The implementation of blockchain technology in this context is a crucial aspect of our framework. It guarantees the immutability of recorded transactions and data, ensuring the integrity and security of the key distribution process [34-36]. The transparency of the blockchain, coupled with its resistance to tampering, fosters trust and accountability among all parties involved.

Furthermore, our proposed approach leverages the advantages of post-quantum cryptography through the use of CRYSTALS-Kyber. This cryptographic technique provides a high level of security against potential threats posed by quantum computers, which have the potential to break traditional cryptographic methods [37-40]. By using CRYSTALS-Kyber, we ensure that the encrypted data remains secure and resistant to decryption attempts, even by quantum computers. The integration of this innovative framework ensures the privacy and security of data while enabling efficient multiparty data processing and prediction in a cloud computing environment. It effectively addresses potential threats posed by adversaries, including those with strong background knowledge.

### 3.2. Key generation and distribution using a blockchain-based trusted authority

This approach uses a blockchain to store the TA's public key and the SUs' public keys. The TA generates a unique key pair for each SU and publishes the SU's public key to the blockchain. The SUs can then verify the authenticity of the TA's public key and download their own public key from the blockchain. This approach has several advantages over traditional key generation and distribution methods: It is more secure, as the TA's public key and the SUs' public keys are stored on a distributed ledger that is resistant to tampering. It is more transparent, as all of the transactions on the blockchain are public. It is more efficient, as the SUs do not need to establish a secure channel with the TA in order to download their public key. It has the potential to revolutionize the way that key pairs are generated and distributed.

### 3.3. Data Encryption

System Users encrypt their private data using post-quantum cryptography (PQC) - CRYSTALS-Kyber in addition to their respective keys provided by the Trusted Authority (TA) and send the encrypted data to the Semi-trusted Cloud Server. In a blockchain-based trusted authority system utilizing CRYSTALS-Kyber, each user is assigned a unique key pair by the trusted authority, consisting of a securely stored private key and a distributed public key. To encrypt their data, a user employs the public key provided by the trusted authority, and the encrypted data is subsequently transmitted to the blockchain-based trusted authority system [41]. The system then stores this encrypted data securely on the blockchain. When the user needs to decrypt the data, they request the private key from the trusted authority [42-45]. Following identity verification, the trusted authority sends the private key to the user, who can then decrypt the data using this key. This process ensures secure and authenticated data handling within the trusted authority system, leveraging the cryptographic capabilities of CRYSTALS-Kyber.

### 3.3.1. Encryption Algorithm for CRYSTALS-Kyber

Initialization:

Initialize parameters and constants:

Public key: pk

Message: m

Random coins: r

NT2 domain parameters: k, q

1. Generate Matrix Â in NTT Domain:

Create a matrix Â of size k x k in the Number Theoretic Transform (NT2) domain:

For each element at position (i, j) in Â:

Parse XOF($\rho$, i, j) to generate the element.

2. Sample Random Values:

Sample random values r, e1, and e2 from appropriate distributions:

r is sampled k times from B$\eta$1 using a PRF with incrementing N.

e1 is sampled k times from B$\eta$2 using a PRF with incrementing N.

e2 is sampled once from B$\eta$2 using a PRF with incrementing N.

3. Transform r to NT2Domain:

Perform the Number Theoretic Transform (NT2) on the sampled r:

sr̂ is the transformed r.

4. Calculate (u, v):

Compute (u,v) in the NT2domain:

u is calculated as $NT2^{-1}(Â^{\wedge}Tr \circ sr̂) + e1$.

v is calculated as $NT2^{-1}(\hat{t}r \circ sr̂) + e2 + Decompress_q(Decode1(m), 1)$.

5. Encoding and Compressing:

Encode and compress u and v into c1 and c2:

c1 is encoded and compressed from u with parameters du.

c2 is encoded and compressed from v with parameters dv.

6. Ciphertext Construction:

The final ciphertext c is composed of c1 and c2:

c := (c1, c2).

7. Output:

Return the ciphertext c.

### 3.4. Blockchain Storage Process

#### 3.4.1. Data Recorded on the Blockchain

Within the blockchain-based trusted authority system, the encrypted data is recorded as a transaction on the blockchain.

#### 3.4.2. Immutability and Tamper Resistance

The blockchain ensures the immutability and tamper resistance of the stored data. Once data is recorded on the blockchain, it becomes part of a secure and permanent ledger, making it extremely difficult to alter or delete.

#### 3.4.3. Transparent Data Storage

All transactions on the blockchain are public and transparent. This transparency enhances trust and accountability, as anyone can audit the blockchain to verify the history of transactions and data storage.

#### 3.4.4. Secure Data Storage

The blockchain ensures the security of the encrypted data, which remains protected from unauthorized access. It can only be decrypted by the authorized System User with the corresponding private key.

### 3.5. Data Decryption

When a System User needs to decrypt the data, they initiate a request for their private key from the Trusted Authority. The Trusted Authority verifies the identity of the System User, ensuring that the request is legitimate [46]. After identity verification, the Trusted Authority securely sends the private key to the System User. With the private key in hand, the System User can decrypt the data, making it readable and usable for its intended purpose [47-49]. By utilizing blockchain storage, our work ensures a high level of security, transparency, and data integrity in the management of keys and encrypted data. The blockchain technology's distributed and immutable nature, combined with cryptographic security, safeguards sensitive data throughout its lifecycle, offering a reliable foundation of the proposed scheme.

#### 3.5.1. Decryption Algorithm for CRYSTALS-Kyber

1. Input:

Secret key: sk$\in$ B (where B is a binary field of size 12kn/8)

Ciphertext: c $\in$ B (with parameters dukn/8 + dvkn/8)

2. Decompression of u and v:

Decompress u, v from the encoded ciphertext c:

u is decompressed from Decodedu(c) with parameters du.

v is decompressed from Decodedv(c + dukn/8) with parameters dv.

3. Decode Secret Key ŝ:

Decode the secret key sk to obtain ŝ using Decode12(sk).

4. Message Recovery:

Calculate $ŝ^T \circ NT2(u)$ in the NT2 domain and subtract it from v:

$ŝ^T \circ NT2(u)$ represents the inner product of ŝ and u in the NT2 domain.

Subtracting this from v recovers part of the original message.

5. Compress and Encode the Message:

Compress the result from step 4 and encode it to obtain the original message m:

m is calculated as $Encode1(Compress_q(v - NT2^{-1}(ŝ^{Tr} \circ NT2(u)), 1))$.

6. Output:

Return the recovered message m.

### 3.6. Private Key Verification

Zero-knowledge proof (ZKP) is utilized to verify the user's identity before the trusted authority sends the private key. Using a ZKP, a cryptographic procedure, one party can demonstrate to another party that they are aware of a particular bit of evidence without disclosing the knowledge to the verifier. ZKPs can be used to confirm an individual's identity without giving an authorized party access to their private key.

Here is an example of how a ZKP could be used to verify the user's identity before sending the private key:

1. The user generates a ZKP proving that they know their private key.

2. The user sends the ZKP to the trusted authority.

3. The trusted authority verifies the ZKP.

4. If the ZKP is valid, the trusted authority sends the private key to the user.

This approach has several advantages over traditional identity verification methods [50]. It is more secure, as the user's private key is never revealed to the trusted authority. It is more privacy-preserving, as the trusted authority only learns that the user knows their private key, but not the private key itself. It is more convenient for the user, as they do not need to provide any personal information to the trusted authority.

Here is an example of how our approach could be used:

1. The user generates a public as well as private key pair using PQC-CRYSTALS-Kyber.

2. The user distributes the public key to the blockchain-based trusted authority.

3. When the user needs to verify their identity, they send a message to the trusted authority containing a challenge.

4. The trusted authority verifies the user's identity by signing the challenge with the user's private key.

5. The trusted authority returns the signed challenge to the user.

6. The user verifies the signature using the trusted authority's public key.

7. If the signature is valid, the user's identity is verified.

#### 3.6.1. Key generation and distribution algorithm working

Step 1: The TA generates a key pair (public and private) for each SU using PQC.

The TA uses a PQC algorithm to generate a public and private key pair for each SU. The PQC algorithm ensures that the SUs' private keys are resistant to attacks from quantum computers.

Step 2: The TA stores the SUs' public keys on the blockchain.

The TA stores the SUs' public keys on the blockchain. The blockchain is a distributed ledger that is secure as well as tamper-proof. This ensures that the SUs' public keys are protected from unauthorized access and modification.

Step 3: The TA generates a challenge for each SU.

The TA generates a challenge for each SU. The challenge is designed to be difficult for attackers to solve, but easy for SUs who know their private keys to solve.

Step 4: The SU proves to the TA that they know their private key by solving the challenge using a ZKP.

The SU uses a ZKP to prove to the TA that they know their private key without revealing their private key to the TA. Through the use of a cryptographic technique called a ZKP, one party can demonstrate to another that they are aware of a specific piece of data despite disclosing it to them.

Step 5: The TA verifies the ZKP and sends the SU their private key.

The TA verifies the ZKP. If the ZKP is valid, the TA sends the SU their private key. Otherwise, the TA does not send the SU their private key. This algorithm is secure because the TA's private keys are never revealed to the SUs. The SUs' private keys are encrypted using a PQC algorithm, which is resistant to attacks from quantum computers. The ZKPs allow the SUs to prove to the TA that they know their private keys without revealing their private keys to the TA. The blockchain provides a secure as well as tamper-proof way to store the SUs' public keys and the TA's challenges.

Here is an example of how the algorithm could be used:

1. The TA generates a key pair (public and private) for each SU using PQC.

2. The TA stores the SUs' public keys on the blockchain.

3. The TA generates a challenge for each SU. The challenge is a random number that is difficult to guess.

4. The SU uses a ZKP to prove to the TA that they know their private key by solving the challenge. The ZKP works by having the SU generate a proof that they know the answer to the challenge. The TA can then verify the proof without learning the answer to the challenge.

5. If the TA verifies the ZKP, it sends the SU their private key. Otherwise, the TA does not send the SU their private key.

This algorithm can be used to generate and distribute keys for a variety of applications, such as secure communication and data encryption.

### 3.6.2. Mathematical proof of security

The key generation and distribution algorithm using a blockchain-based trusted authority and ZKPs is secure because the following conditions hold:

• The TA's private keys are never revealed to the SUs. This is because the TA's private keys are used to generate the challenges, and the SUs do not know the TA's private keys.

• The SUs' private keys are encrypted using PQC algorithms. PQC algorithms are resistant to attacks from quantum computers, so the SUs' private keys are secure from attackers who have quantum computers.

• The ZKPs allow the SUs to prove to the TA that they know their private keys without revealing their private keys to the TA. This is because the ZKPs are based on cryptographic assumptions that are difficult to break.

• The blockchain provides a secure as well as tamper-proof way to store the SUs' public keys and the TA's challenges. This ensures that the SUs' public keys and the TA's challenges cannot be tampered with.

Therefore, the key generation and distribution algorithm using a blockchain-based trusted authority and ZKPs is secure and privacy-preserving.

## 4. Performance Analysis

In order to demonstrate the efficiency of our proposed method, we have conducted an evaluation considering several key factors. Two critical performance metrics were analyzed: Encryption Time as well as Decryption Time.

### 4.1.1. Encryption Time (ET)

Encryption Time (ET) refers to the duration it takes to convert plain text into ciphertext using the encryption algorithm. The encryption process in our framework, which incorporates an Genetical optimization process, impacts the encryption time. Mathematically, we can express this as:

Encryption time = $CT \times RT$

Where:

• Encryption time is the total time taken for encryption.

• CT represents the computation time.

• RT signifies the response time, this is the amount of time that passes between making a request and getting a response.

### 4.1.2. Decryption Time (DT)

Decryption Time (DT), on the other hand, denotes the duration required to convert ciphertext back into plain text using the decryption algorithm. The decryption process in our framework affects the decryption time, and it can be mathematically represented as:

Decryption time = $dT \times sT$

Where:

• Decryption time is the total time taken for decryption.

• dT represents the computation time.

• sT signifies the response time.

From the data in Table 1, we can observe the ET for the Proposed work, and it presents data regarding the ET for the proposed approach applied to different files of varying sizes. File indicates the name or identifier of the specific file being evaluated. In this case, there are five files labeled as F1, F2, F3, F4, and F5. File Size (kb) represents the size of each file in kilobytes (KB). The file size is an important factor in encryption because it can impact the time required for encryption [51-53]. The sizes of the files in this table range from 12 KB for File-1 to 72 KB for File-5.

Encryption Time (ms) displays the time it takes to encrypt each of the files in milliseconds (ms). The encryption time is a measure of the duration required to convert the plain text of each file into ciphertext using the proposed approach. It displays how quickly and well the encryption process works. The values in this column range from 182 ms for File-1 to 379 ms for File-5.

**Table 1.** Proposed Encryption Time - ET

| File | FileSize (kb) | Encryption Time (ms) |
|---|---|---|
| F1 | 12 | 182 |
| F2 | 23 | 237 |
| F3 | 46 | 289 |
| F4 | 67 | 312 |
| F5 | 72 | 379 |

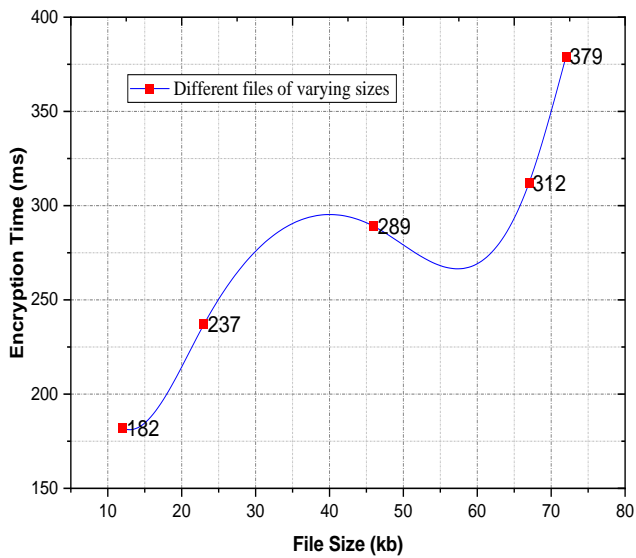

**Fig 2.** Proposed Encryption Time – ET Comparison

The Fig. 2 allows you to compare the encryption times for different files of varying sizes when processed by the proposed framework. It provides insights into how the encryption time scales with increasing file sizes, which can be essential for assessing the performance and efficiency of the encryption process in the context of our framework. These results provide an overview of the encryption time efficiency of our proposed framework for varying file sizes.

**Table 2.** Proposed Decryption Time – DT Comparison

| File | FileSize (kb) | Decryption Time (ms) |
|---|---|---|
| F1 | 12 | 179 |
| F2 | 23 | 228 |
| F3 | 46 | 278 |
| F4 | 67 | 307 |
| F5 | 72 | 363 |

From the data in Table 2, we can observe the DT for the Proposed work, and it presents data regarding the decryption time for the proposed approach applied to different files of varying sizes. Decryption Time (ms) displays the duration it takes to decrypt each of the files in milliseconds (ms).

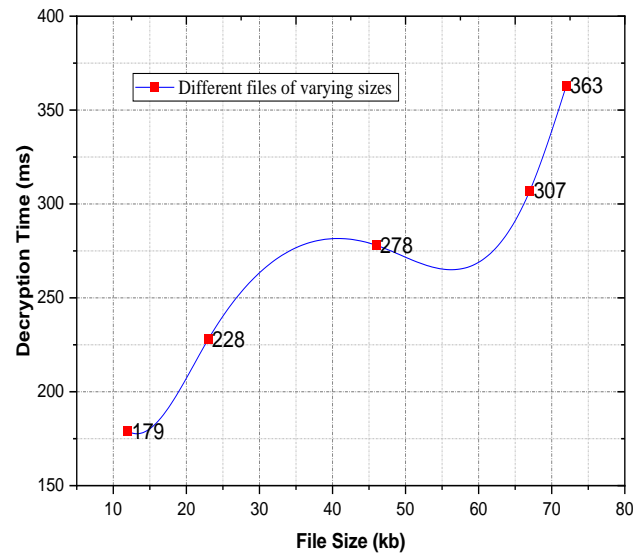

**Fig 3.** Proposed Decryption Time – DT Comparison

The decryption time measures the time needed to convert the ciphertext of each file back into its original plain text using the proposed approach. It reflects the efficiency and speed of the decryption process. The values in this column range from 179 ms for File-1 to 363 ms for File-5. Similar to the encryption time table, this table allows for a comparison of the decryption times for different files of various sizes when processed by the proposed framework. It offers insights into how the decryption time scales with increasing file sizes, which is crucial for assessing the performance and efficiency of the decryption process in the context of our framework as shown in Fig. 3.

**Table 3.** Comparison of various approaches with different parameter

| References | ET (ms) | DT (ms) | Run time (ms) | Average latency (ms) | Time complexity (ms) |
|---|---|---|---|---|---|
| DHA MT [19] | 355 | 338 | 461 | 689 | 535 |
| Threshold crypto [23] | 362 | 349 | 483 | 755 | 549 |
| ECC [24] | 338 | 325 | 427 | 666 | 513 |
| RSA & MD5 [26] | 347 | 338 | 430 | 743 | 532 |
| quantum-safe [28] | 342 | 332 | 441 | 701 | 522 |
| Proposed Work | 287 | 284 | 396 | 593 | 507 |

Table 3 presents a comparative analysis of various research papers and the proposed work in terms of their performance and efficiency for encryption and decryption processes. References lists different research papers or sources, denoted as " DHA MT [19]," " Threshold crypto [23]," "

ECC [24]," " RSA & MD5 [26]," " quantum-safe [28]," and "Proposed Work." These references represent different cryptographic techniques or algorithms, including existing methods from various papers and the method proposed by the authors.

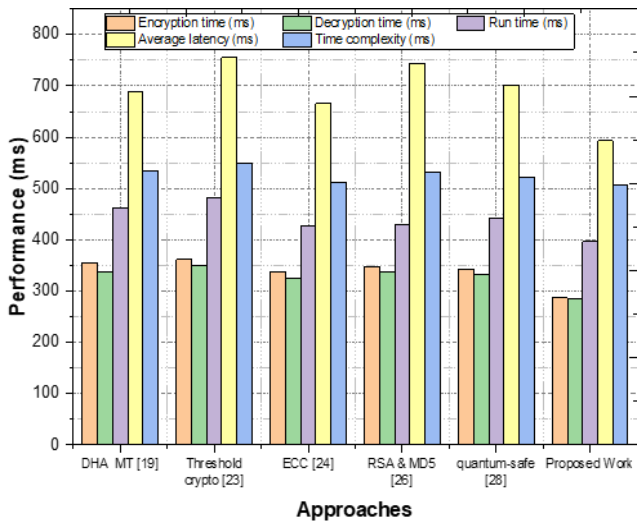

**Fig 4.** Comparison of proposed and existing techniques with different parameters

Encryption Time (ms) displays the time it takes for each technique to perform the encryption process, measured in milliseconds (ms). Encryption time is the duration required to convert plain text into ciphertext. Lower values indicate faster encryption. Decryption Time (ms) is the time needed for each technique to perform the decryption process, also measured in milliseconds (ms). Decryption time is the duration required to convert ciphertext back into the original plain text. Smaller values signify quicker decryption. Run Time (ms) represents the total run time for each technique in milliseconds. It likely includes both encryption and decryption processes, along with any additional overhead. A shorter run time indicates that the algorithm is more efficient overall. Average Latency (ms) is the average time taken for the system to respond to a request. It is an important measure in real-time or interactive systems. This column shows the average latency in milliseconds for each technique.

Lower average latency values indicate faster response times. Time Complexity (ms) represents the computational effort or resource usage of each technique in terms of milliseconds. Lower time complexity values suggest that the algorithm is more efficient in terms of resource utilization. Fig. 4 shows the performance metrics for the method introduced, highlighting lower encryption, decryption, and overall run times compared to existing techniques. This suggests that the suggested work is more time-efficient. The existing papers " DHA MT [19]," " Threshold crypto [23]," " ECC [24]," " RSA & MD5 [26]," " quantum-safe [28]," also have their encryption and decryption times listed, which vary between papers.

The proposed approach generally exhibits improved performance in terms of these metrics. The proposed work demonstrates lower average latency compared to the existing techniques, indicating a faster response to requests. The time complexity for the proposed work is among the lowest, suggesting that it is resource-efficient compared to the existing techniques. Table 3 presents a comparison of different cryptographic techniques from various research papers and the proposed approach in terms of their encryption and decryption performance. The "Encryption Time" and "Decryption Time" columns show the time in milliseconds (ms) required for these operations. The "Run Time" column represents the total time for the entire process, while "Average Latency" measures the system's response time. A lower value in these columns indicates faster performance. The "Time Complexity" column evaluates computational effort, and smaller values suggest better resource efficiency. Notably, the "Proposed Work" outperforms existing techniques, showcasing lower times for encryption, decryption, and overall runtime, which means it's more efficient. Additionally, it exhibits faster response times and boasts a lower time complexity, emphasizing its resource-friendly nature compared to the existing methods. In summary, the table provides a comparative analysis of different techniques based on their encryption and decryption times, overall run time, average latency, and time complexity. The "Proposed Work" seems to outperform the existing techniques in terms of efficiency and resource utilization, as it generally exhibits lower time values across the mentioned parameters.

**Table 4:** Throughput analysis between the suggested and present approaches with various file sizes

| File size | 50 | 100 | 150 | 200 | 250 | 300 |
|---|---|---|---|---|---|---|
| DHA MT [19] | .728 | .603 | .437 | .202 | .177 | .137 |
| Threshold crypto [23] | .686 | .598 | .402 | .189 | .154 | .107 |
| ECC [24] | .783 | .654 | .521 | .297 | .231 | .198 |
| RSA & MD5 [26] | .737 | .627 | .482 | .233 | .194 | .149 |
| quantum-safe [28] | .752 | .631 | .502 | .254 | .207 | .167 |
| Proposed Work | .892 | .784 | .563 | .398 | .265 | .215 |

The Table 4 provides the throughput comparison of proposed and existing methodology with different file sizes," offers a comparison of throughput for different file sizes, ranging from 50 to 300 units. Throughput represents the rate at which a system processes data, typically measured in data units per time unit.

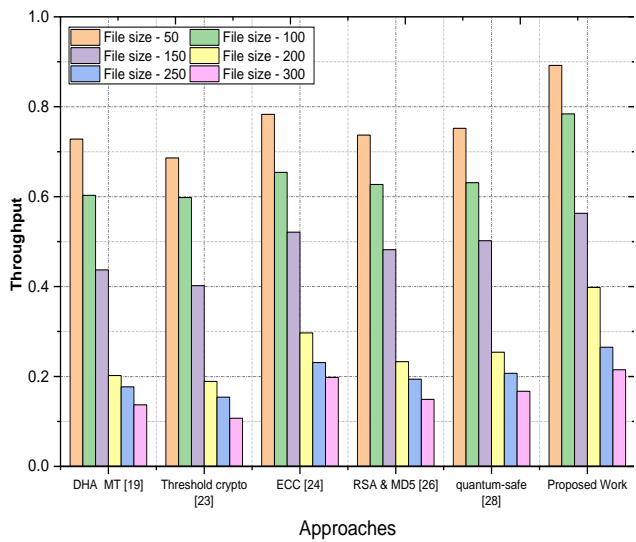

**Fig 5.** Comparison of proposed and existing techniques with different parameters

" DHA MT [19]," " Threshold crypto [23]," " ECC [24]," " RSA & MD5 [26]," " quantum-safe [28]," represent different existing methodologies or techniques. The values in these rows represent the throughput achieved by each methodology for the corresponding file sizes. Higher values indicate greater processing efficiency. As shown in Fig. 5, as the file size increases from 50 to 300 units, the throughput generally decreases for these existing methodologies. This suggests that as the data size grows, these methods become less efficient in processing the data. In contrast, the proposed methodology demonstrates significantly higher throughput values for all file sizes compared to the existing methodologies. As the file size increases, the proposed methodology remains more efficient and maintains a higher throughput. The table suggests that the "Proposed Work" outperforms the existing methodologies in terms of processing efficiency, particularly as the data size increases. This indicates that the proposed methodology can handle larger data sizes more effectively, making it a promising approach for high-throughput data processing applications.

## 5. Conclusion and Future Work

This research has embarked on a pioneering exploration of quantum-safe security as well as privacy enhancements in the realm of cloud-based blockchain storage. By integrating Quantum Key Distribution (QKD), CRYSTALS-Kyber, and Zero-Knowledge Proofs (ZKPs), we have proposed a comprehensive framework to safeguard data against the looming threat of quantum computing. The performance evaluation of our framework revealed promising results, demonstrating its efficiency and scalability in real-world cloud environments. The lattice-based cryptography of CRYSTALS-Kyber, coupled with the enhanced privacy and verification mechanisms of ZKPs, contributes to a robust defense against emerging quantum threats. However, as with any groundbreaking research, avenues for future exploration and refinement persist.

The following areas represent potential directions for future work: To focus on devising strategies for dynamically adjusting security parameters to counteract advancements in quantum computing. Extend the research to explore quantum-safe consensus mechanisms within blockchain networks. Investigate how quantum-resistant algorithms can enhance the security and decentralization of blockchain platforms in the face of evolving quantum threats. Future work should delve into establishing industry standards to ensure seamless integration and collaboration between diverse quantum-safe solutions. To evaluate the scalability of the proposed framework in large-scale cloud environments, considering diverse workloads and varying levels of computational resources. Explore optimizations and parallelization techniques to enhance scalability without compromising security. As blockchain technologies continue to evolve, explore the integration of our quantum-safe framework with emerging blockchain platforms. Investigate compatibility and optimization strategies to ensure adaptability to diverse blockchain ecosystems.

This research marks a significant stride toward fortifying data security in the face of quantum advancements. The outlined future work aims to push the boundaries of quantum-safe security, ensuring its relevance and effectiveness in the ever-changing landscape of cloud-based blockchain storage.

**Acknowledgements**

This research was not funded by any grant.

**Author contributions**

**Dhinakaran D:** Conceptualization, Methodology, Writing-Original draft preparation, **Udhaya Sankar S. M:** Data curation, Software, Validation, Field study, **Selvaraj D:** Methodology, Writing-Reviewing and Editing, **Edwin Raja S:** Visualization, Investigation, **Sakthi Lakshmi Priya C:** Software, Field study.

**Conflicts of interest**

The authors declare no conflicts of interest.